\documentclass[11pt,aps,pra]{revtex4-1}
\usepackage{amsmath}    
\usepackage{graphicx}   

\def\bra#1{\langle\, #1\, |}
\def\ket#1{|\, #1\, \rangle}

\begin{document}
\title{Dissociative recombination by frame transformation to Siegert pseudostates: A~comparison with
a numerically solvable model}
\author{D\'{a}vid Hvizdo\v{s}}
\affiliation{J. Heyrovsk\'{y} Institute of Physical Chemistry, ASCR,
Dolej\v{s}kova 3, 18223 Prague, Czech Republic}
\affiliation{Institute of Theoretical Physics, Faculty of Mathematics and Physics, Charles University in Prague, V Hole\v{s}vi\v{c}k\'{a}ch 2, 180 00 Prague, Czech Republic}
\author{Martin V\'{a}\v{n}a}
\affiliation{Institute of Theoretical Physics, Faculty of Mathematics and Physics, Charles University in Prague, V Hole\v{s}vi\v{c}k\'{a}ch 2, 180 00 Prague, Czech Republic}
\author{Karel Houfek}
\affiliation{Institute of Theoretical Physics, Faculty of Mathematics and Physics, Charles University in Prague, V Hole\v{s}vi\v{c}k\'{a}ch 2, 180 00 Prague, Czech Republic}
\author{Chris H.~Greene}
\affiliation{Department of Physics and Astronomy, Purdue University, West Lafayette,
Indiana 47907, USA}
\author{Thomas N. Rescigno}
\affiliation{Chemical Sciences Division, Lawrence Berkeley National Laboratory, Berkeley,
California 94720, USA}
\author{C.~William McCurdy}
\affiliation{Chemical Sciences Division, Lawrence Berkeley National Laboratory, Berkeley,
California 94720, USA}
\affiliation{Department of Chemistry, University of California, Davis, California 95616, USA}
\author{Roman \v{C}ur\'{\i}k}
\affiliation{J. Heyrovsk\'{y} Institute of Physical Chemistry, ASCR,
Dolej\v{s}kova 3, 18223 Prague, Czech Republic}
\date{\today}

\begin{abstract}
\vspace{-4mm}
We present a simple two-dimensional model of the indirect dissociative recombination process.
The model has one electronic and one nuclear degree of freedom and it can be solved to high
precision, without making any physically motivated approximations, by employing
the exterior complex scaling method together with the finite-elements method and
discrete variable representation. The approach is applied to solve a model
for dissociative recombination of H$_2^+$ and the results serve as a benchmark
to test validity of several physical approximations commonly used in the
computational modeling of dissociative recombination for real molecular targets.
The second, approximate, set of calculations employs a combination of multi-channel
quantum defect theory and frame transformation into a basis of Siegert pseudostates.
The cross sections computed with the two methods are compared in detail for collision energies
from 0 to 2 eV.
\end{abstract}

\maketitle

\section{\label{sec-intro}Introduction}

Dissociative recombination, one of the most fundamental electron-induced chemical
rearrangement processes, is important for understanding the chemical dynamics of
interstellar clouds, as well as the chain of reactive processes in low temperature
plasmas \cite{Larsson_Orel_DRbook_2008}.

The frame transformation (FT) technique \cite{Chang_Fano_1972} is a well-established
procedure to model rovibrationally inelastic collisions of electrons with neutral molecules and
cations. Moreover, there has been a number of studies that successfully employed this
technique for the dissociative recombination (DR) process
\begin{equation}
\label{eq-DR}
\textrm{e}^- + \textrm{AB}^+ \rightarrow \textrm{A} + \textrm{B}.
\end{equation}
The molecular cations in these studies were
H$_2^+$ \cite{Hamilton_Greene_PRL_2002}, H$_3^+$ \cite{Kokoouline_Greene_PRA_2003},
LiH$^+$ \cite{Curik_Greene_PRL_2007,Curik_Greene_MP_2007},
NO$_2^+$ \cite{Haxton_Greene_NO2_PRA_2012},
LiHe$^+$ \cite{Curik_Gianturco_2013}, LiH$_2^+$ \cite{Haxton_Greene_LiH2_PRA_2008},
and HeH$^+$ \cite{Haxton_Greene_HeH_2009,Curik_Greene_JCP_2017}.
All of these calculations, except the LiH$_2^+$ and HeH$^+$ cases, used Siegert pseudostates
\cite{Tolst_sieg_1997,Tolst_sieg_1998} for the nuclear vibrational basis and
they all exploited the following two-step procedure:
\begin{itemize}
\item[1.] The fixed-nuclei $S$ matrix is frame-transformed into a subset of nuclear Siegert
pseudostates. This subset contains real-valued bound states and complex-valued outgoing-wave states
that discretize the nuclear continuum. Stability of the results is typically tested for various
sizes of the nuclear box and number of the continuum states. The frame transformation formula
comes as a modification of the well-known FT expression \cite{Chang_Fano_1972} with addition
of a surface term as
\begin{equation}
\label{eq-Smat-FT}
S_{jj'}= \int_0^a dR \phi_j(R) S(R)\phi_{j'}(R) +
i\frac{\phi_j(a)S(a)\phi_{j'}(a)}{K_j+K_{j'}}.
\end{equation}
This formula appears for the first time in work of Hamilton and Greene
\cite{Hamilton_Greene_PRL_2002} in 2002 and in Hamilton's PhD thesis
\cite{Hamilton_thesis}. While it has a correct limit
for small quantum defects (giving orthogonality of the Siegert pseudostates),
its derivation has never been explained nor published. It has been called an \emph{ad hoc} formula
in \cite{Haxton_Greene_HeH_2009} where its validity was questioned. Nonetheless,
the expression (\ref{eq-Smat-FT}) is one of the cornerstones for most of the studies listed
above.
\item[2.] After the elimination of the closed electronic channels the DR rate is computed in a form
of the missing electronic flux. The physical $S$ matrix $S^{phys}$ appears to be subunitary
in the nuclear basis of the bound and outgoing-wave Siegert pseudostates.
This means that the electronic
flux is being lost during the collision. Hamilton and Greene \cite{Hamilton_Greene_PRL_2002}
realized that the only way to lose the electronic flux is through electronic
recombination and following dissociation.

All the studies listed above used this physical reasoning and the DR cross section for the initial
state labeled with $j'$ was calculated as
\begin{equation}
\sigma_{j'} = \frac{\pi}{2(E-\epsilon_{j'})} \left[ 1 -
\sum_j S^{phys}_{jj'}(E) S^{phys\,\dag}_{j'j}(E) \right],
\end{equation}
where $E$ is the total energy and $\epsilon_{j'}$ is the initial channel energy.
\end{itemize}

While this two-step computational strategy produced the state-of-the-art theoretical
DR cross sections data,
it is clear that it contains two major theoretical leaps that require physical explanation or/and
well-controlled numerical evidence. Therefore, the goal of the present study is to provide
firmer theoretical grounds for the expression (\ref{eq-Smat-FT}) and to convince the reader
that the ideas behind the second step are indeed correct. For the latter we choose to provide
numerical evidence by comparing the results with benchmark data
which are obtained with a numerically solvable model of H$_2^+$ in two dimensions,
with one electronic and one nuclear coordinate.
The model is similar to the two-dimensional model of resonant
electron-molecule collisions, which was introduced in \cite{Houfek_Rescigno_McCurdy_PRA_2006} and used
to test the local and nonlocal theory of nuclear dynamics of these collisions
\cite{Houfek_Rescigno_McCurdy_PRA_2006,Houfek_Rescigno_McCurdy_PRA_2008}. The numerical technique used
to solve this model is based on the exterior complex scaling approach combined with the finite-elements
method and discrete variable representation
\cite{Rescigno_McCurdy_PRA_2000,McCurdy_Baertschy_Rescigno_JPB_2004}
and the same numerical approach is also used in the present study with limitations described below.

Adaptation of this 2D model of vibrational excitation and dissociative electron attachment for
dissociative recombination will be described in the following section.
In Sec.~\ref{sec-ft} we demonstrate how the FT formula (\ref{eq-Smat-FT}) can be derived
more rigorously and we also describe the multi-channel quantum defect (MQDT) procedure
applied to the present model system. The DR cross sections are compared in detail
in Sec.~\ref{sec-res} for the collision energy range 0--2 eV. In Sec.~\ref{sec-con}
we comment on the differences between the two approaches and
we also discuss possible improvement and outlook for
the FT procedure. Finally, the mathematical detail of the expansion in Siegert basis pseudostates
in our derivation of Eq.~(\ref{eq-Smat-FT}) are given in the Appendix.

If not stated otherwise, all relations and values in tables are in atomic units,
in which $\hbar = m_e = e = 4\pi\epsilon_0 = 1$. Internuclear distances are
given in units of the Bohr radius $ 5.291\,772\times 10^{-11}\ \textrm{m}$,
cross sections in units of $\textrm{Bohr}^2 = 2.800\,285\times 10^{-21}\ \textrm{m}^2$.

\section{\label{sec-exmod}Numerically solvable H$_2^+$-like model}

\subsection{\label{ssec-theory}Theoretical description}

The model Hamiltonian employed in the present paper is
\begin{equation}
\label{eq-Ham-tot}
        H(R,r) = H_0(R,r) + V(R,r) = H_N(R) + H_e(r) + V(R,r),
\end{equation}
with
\begin{eqnarray}
\label{eq-Ham-twoN}
H_N &=& -\frac{1}{2 M}\frac{\partial^2}{\partial R^2} + V_0(R), \\
\label{eq-Ham-twoE}
H_e &=& - \frac{1}{2}\frac{\partial^2}{\partial r^2} -
\frac{1}{r} + \frac{l(l+1)}{2r^2},
\end{eqnarray}
where $V_0(R)$ is the ground state potential curve of the ${}^2\Sigma_g^+$ state of the H$_2^+$ ion, approximated by the Morse potential in the present study
\begin{equation}
\label{eq-V0}
  V_0(R) = D_0 \left(e^{-2\alpha_0(R-R_e)}-2\,e^{-\alpha_0(R-R_e)}\right)\;,
\end{equation}
with $D_0 = 0.1027$ Hartree, $\alpha_0 = 0.69$ Bohr$^{-1}$, $R_e = 2.0$ Bohrs.
The symbol $M = 918.076$ a.u. denotes the reduced mass of H$_2^+$, and
$l$ is the angular momentum of the incoming electron. The interaction $V(R,r)$  coupling the
electronic and nuclear degrees of freedom is taken from Edward Hamilton's PhD thesis
\cite{Hamilton_thesis}
\begin{equation}
\label{eq-Ham-pot}
V(R,r) = -\alpha_1 \left(1 - \tanh\frac{\alpha_2-R-\alpha_3 R^4}{7}\right)
\tanh^4 \left(\frac{R}{\alpha_4}\right) \frac{e^{-r^2/3}}{r},
\end{equation}
where $\alpha_1 = 1.6435$, $\alpha_2 = 6.2$, $\alpha_3 = 0.0125$, and $\alpha_4 = 1.15$.
The form of the potential in Eq.~(\ref{eq-Ham-pot}) taken together with the potentials
in Eqs.~(\ref{eq-Ham-twoN}) and (\ref{eq-Ham-twoE}) is designed to mimic the
${}^1\Sigma_u$ Rydberg states of H$_2$.

The total wave function $\psi_{E}^+(R,r)$ satisfying the Schr\"{o}dinger equation
\begin{equation}
\label{eq-Schr-2d}
        (E-H)\psi_{E}^+(R,r) = 0
\end{equation}
can be split into the initial and scattered parts as
\begin{gather}
\label{eq-wfn-split}
        \psi_{E}^+(R,r) = \psi_{in}(R,r) + \psi_{sc}(R,r), \\
        (E-H_0)\psi_{in}(R,r) = 0.
\end{gather}
The scattered part is then a solution of the so-called driven Schr\"{o}dinger equation
\begin{equation}
\label{eq-Schr-scat}
        (E-H)\psi_{sc}(R,r) = V(R,r) \psi_{in}(R,r)
\end{equation}
with the initial state
\begin{equation}
\label{eq-wfn-init}
\psi_{in}(R,r)=\chi_{j'} (R) \Phi_{k_{j'},l}(r),
\end{equation}
constructed from a bound nuclear state $\chi_{j'}(R)$ and a free incoming electronic state
defined by the momentum $k_{j'}$. These states are eigenstates of the following Hamiltonians
\begin{eqnarray}
\label{eq-Ham-2d}
H_N(R) \chi_{j'} (R)&=&\epsilon_{j'} \chi_{j'} (R),\\
H_e(r)\Phi_{k_{j'},l}(r)&=&\frac{k_{j'}^2}{2}\Phi_{k_{j'},l}(r).
\end{eqnarray}
The function $\Phi_{k_{j'},l}$ is thus an energy-normalized spherical Coulomb function and
obviously $E = \epsilon_{j'} + k_{j'}^2/2$.

The asymptotic boundary condition for the scattered wave gives
\begin{equation}
\label{eq-wfn-2d-asy}
\psi_{sc}(R,r) \xrightarrow[R\to\infty]{ } \sqrt{\frac{2}{\pi k_{j'}}} \sum_{n}
f^{\mathrm{DR}}_{j'\to n}\,\rho_{n}(r)\, e^{iK_n R},
\end{equation}
where $f^{\mathrm{DR}}_{j'\to n}$ is the DR scattering amplitude from the initial vibrational state $j'$ to the final Rydberg state $\rho_n$. The Rydberg state function $\rho_{n}(r)$ of the electron satisfies
\begin{equation}
\label{eq-Schr-Ryd}
[H_e + V_\infty(r)]\rho_{n}(r) = E_{n} \rho_{n}(r),
\end{equation}
where $V_\infty(r) = \lim_{R\to\infty}V(R,r)$ and again $E=E_n+\frac{K_n^2}{2 M}$.

In case of dissociative recombination the outgoing states are a product of an unperturbed nuclear continuum state with momentum $K_n$ with zero angular momentum and a bound $n$-th Rydberg state of the electron $\rho_n(r)$ with energy $E_n$
\begin{equation}
\label{eq-wfn-out}
        \psi_{out,n}(R,r)=\sqrt{\frac{2M}{\pi K_n}} \sin(K_n R) \rho_n(r).
\end{equation}
These states are energy-normalized solutions to the Schr\"{o}dinger equation with the DR channel Hamiltonian
\begin{equation}
\label{eq-Ham-asyR}
        H_{\mathrm{DR}}(R,r)=-\frac{1}{2} \frac{\partial^2} {\partial r^2 }-\frac{1}{2M}
\frac{\partial^2}{\partial R^2 }-\frac{1}{r}+\frac{l(l+1)}{2r^2}+V_\infty(r),
\end{equation}
which is the limit of the original full Hamiltonian (\ref{eq-Ham-tot}) for large internuclear distances $R$.

The $T$ matrix for the DR channel is then expressed as
\begin{equation}
\label{eq-DR-Tmat}
        T^{\mathrm{DR}}_{j' \to n}(E) = \bra{\psi_{out,n}} V_{\mathrm{DR}} \ket{\psi_E^+},
\end{equation}
with the channel potential
\begin{equation}
\label{eq-VDR}
        V_{\mathrm{DR}}(R,r) = V(R,r) - V_\infty(r) + V_0(R).
\end{equation}
Finally, the resulting DR cross section for the initial vibrational state $j'$ and the
final Rydberg state $n$ is given by
\begin{equation}
\label{eq-CS-DR}
\sigma^{\mathrm{DR}}_{j' \rightarrow n}(E)=\frac{4\pi^3}{k_{j'}^2}\left |
T^{\mathrm{DR}}_{j' \rightarrow n}(E) \right |^2.
\end{equation}

\subsection{\label{sec-num-imp}Numerical implementation}

The actual solution of Eq.~(\ref{eq-Schr-scat}) was obtained by using a combination of
finite-elements method (FEM), discrete variable representation (DVR),
and exterior complex scaling (ECS) methods \cite{Rescigno_McCurdy_PRA_2000}.
The FEM-DVR method serves to discretize the continuous variables (electronic and nuclear coordinates)
by dividing the assumed region into several finite elements (FEM) and then
creating a basis function set on each element (DVR). Specifically, the DVR basis is made
up of Lagrange interpolation polynomials through Gauss-Lobatto quadrature points on each
finite element (additionally altered by certain boundary conditions).
This basis can then be used to approximate any function (with some degree of accuracy)
on the aforementioned assumed region.

Lastly, the exterior complex scaling (ECS) is a method of bending a coordinate
(let us say $R$) into the complex plane at some point $R_0$.
This point should be far beyond the interaction region. So the new coordinate $R'$ satisfies
\begin{equation}
	R'(R)=\left\{\begin{matrix}
R, & R<R_0,\\
R_0+(R-R_0)e^{i\theta}, & R\geq R_0,
\end{matrix}\right.
\label{eq:ECS}
\end{equation}
where $\theta$ is a bending angle. The main advantage of the ECS approach is that it unifies
the asymptotic boundary conditions for bound and outgoing continuum states. In the present two-dimensional model we employ the ECS method for both the electronic $r$ and nuclear $R$ coordinates.

The final values of parameters of the FEM-DVR grids for the
$e^- + \mathrm{H}_2^+$ DR model are presented in Table~\ref{table:grids}.
They were settled on after extensive tests of convergence. We should note
that the bending angle $\theta$ for the nuclear coordinate $R$ in the presented model must be less than $\pi/8$ and for the electronic coordinate $r$ less than $\pi/4$ to avoid divergence of $V(R,r)$ for large $R$ and $r$, respectively.
The spherical Coulomb functions are evaluated using COULCC routines \cite{COULCC}.

\begin{table}[!ht]
\centering
\begin{tabular}{|c ||c |c |c |c |c |c |c |}
 \hline
 \multicolumn{8}{|c|}{\textbf{Electronic coordinate parametrization, $n_q=6,\quad \theta=20^{\circ}$}} \\
 \hline
 \multicolumn{8}{|l|}{Real part}\\
 \hline
 Endpoints & 1.0 & 4.0 & 20.0 & 100.0 & 1300.0 & \multicolumn{2}{c|}{-} \\
 \hline
 No. of elements & 8 & 12 & 8 & 16 & 120 & \multicolumn{2}{c|}{-} \\
 \hline
 \multicolumn{8}{|l|}{Complex scaled part}  \\
 \hline
 Endpoints & 1350.0 & 1400.0 & 1500.0 & 1700.0 & 2000.0 & 3000.0 & 100000.0 \\
 \hline
 No. of elements & 5 & 2 & 1 & 1 & 1 & 1 & 5 \\
 \hline \hline
 \multicolumn{8}{|c|}{\textbf{Nuclear coordinate parametrization, $n_q=6,\quad \theta=20^{\circ}$}} \\
 \hline
 \multicolumn{8}{|l|}{Real part}\\
 \hline
 Endpoints & 1.0 & 3.0 & 4.0 & 12.0 & \multicolumn{3}{c|}{-}\\
 \hline
 No. of elements & 12 & 24 & 12 & 120 & \multicolumn{3}{c|}{-}\\
 \hline
 \multicolumn{8}{|l|}{Complex scaled part}\\
 \hline
 Endpoints & 12.5 & 14.0 & 18.0 & 58.0 & 200.0 & 1000.0 & 10000.0\\
 \hline
 No. of elements & 8 & 6 & 2 & 4 & 3 & 3 & 3\\
 \hline
\end{tabular}
\caption{
The final values of parameters of the FEM-DVR grids used in calculations. Both grids start at 0.0. The number of elements under each endpoint is the number of finite elements on the interval between the previous and the respective endpoint.
For example, the first real-part interval $[0.0, 1.0]$ is split into 8 elements.
All the listed "Endpoints" values are in atomic units.
The value $n_q$ is the Gauss-Lobatto quadrature order and $\theta$ is the ECS bending angle.}
\label{table:grids}
\end{table}

\begin{figure}[!t]
\includegraphics[width=1\textwidth]{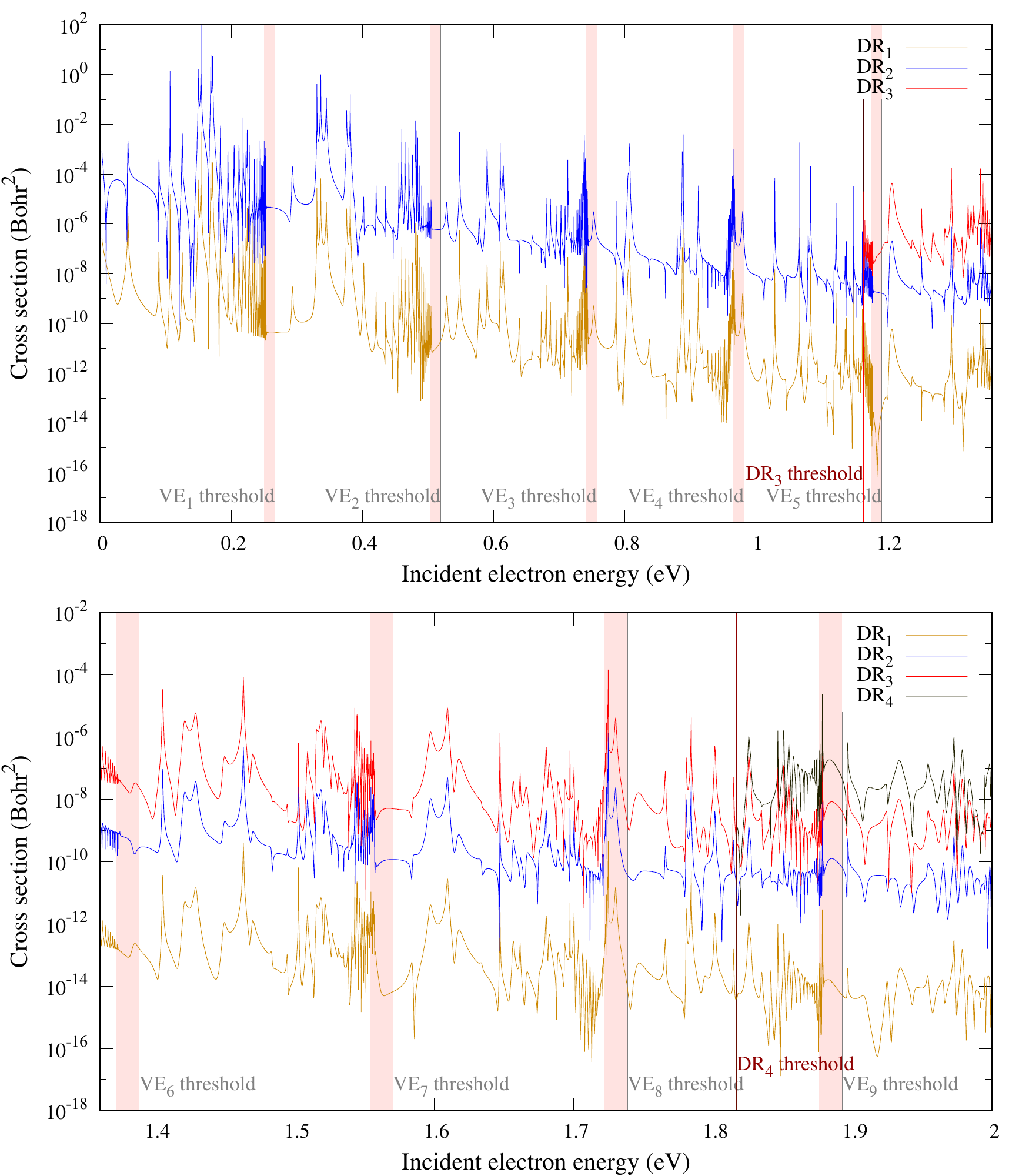}
\caption{\label{fig-DR}
(Color online) The dissociative recombination cross sections of the first four Rydberg channels.
The light pink regions show where the calculated values are not converged (see the text).
DR$_n$ labels the DR cross section $\sigma^{\mathrm{DR}}_{j' \rightarrow n}(E)$
for $n$-th Rydberg channel and initial vibrational state $j'=0$.
}
\end{figure}

The strength of the presented approach lies in the fact that it makes no \textit{physical}
approximations. The only approximations are of a numerical nature (e.g. discretizing continuous
variables). There is however one drawback (aside from the calculations being time consuming) stemming
from the chosen numerical approach. Ideally, when plotting the energy dependence of the cross section
$\sigma^{\mathrm{DR}}_{j' \rightarrow n}(E)$, there would be an infinite number of closed-channel
resonances accumulating just below each vibrational threshold corresponding to an infinite number of
Rydberg states. Any numerical implementation, however, works with a finite range of
the electronic grid, giving only a finite number of these Rydberg states. Therefore, there will always
exist a particular energy window, just below every vibrational excitation
threshold, where the computed cross sections are incorrect and not converged.
The cross section in such a window is dominated by vibrational Feshbach resonances describing
a neutral state in which the molecule is vibrationally excited (to a vibrational state corresponding to
the threshold) and the colliding electron becomes bound in a high-$n$ Rydberg state. One can shrink
these regions by enhancing the maximum electronic grid distance,
but it is impossible to remove these energy windows completely.
These shortcomings are demonstrated as light pink energy windows in Fig.~\ref{fig-DR}
displaying the computed DR cross section as a function of the collision energy up to 2 eV.
The number of Rydberg states which are well represented on the final grid given in
Table~\ref{table:grids} is about 20.
\begin{figure}[tbh]
\includegraphics[width=1\textwidth]{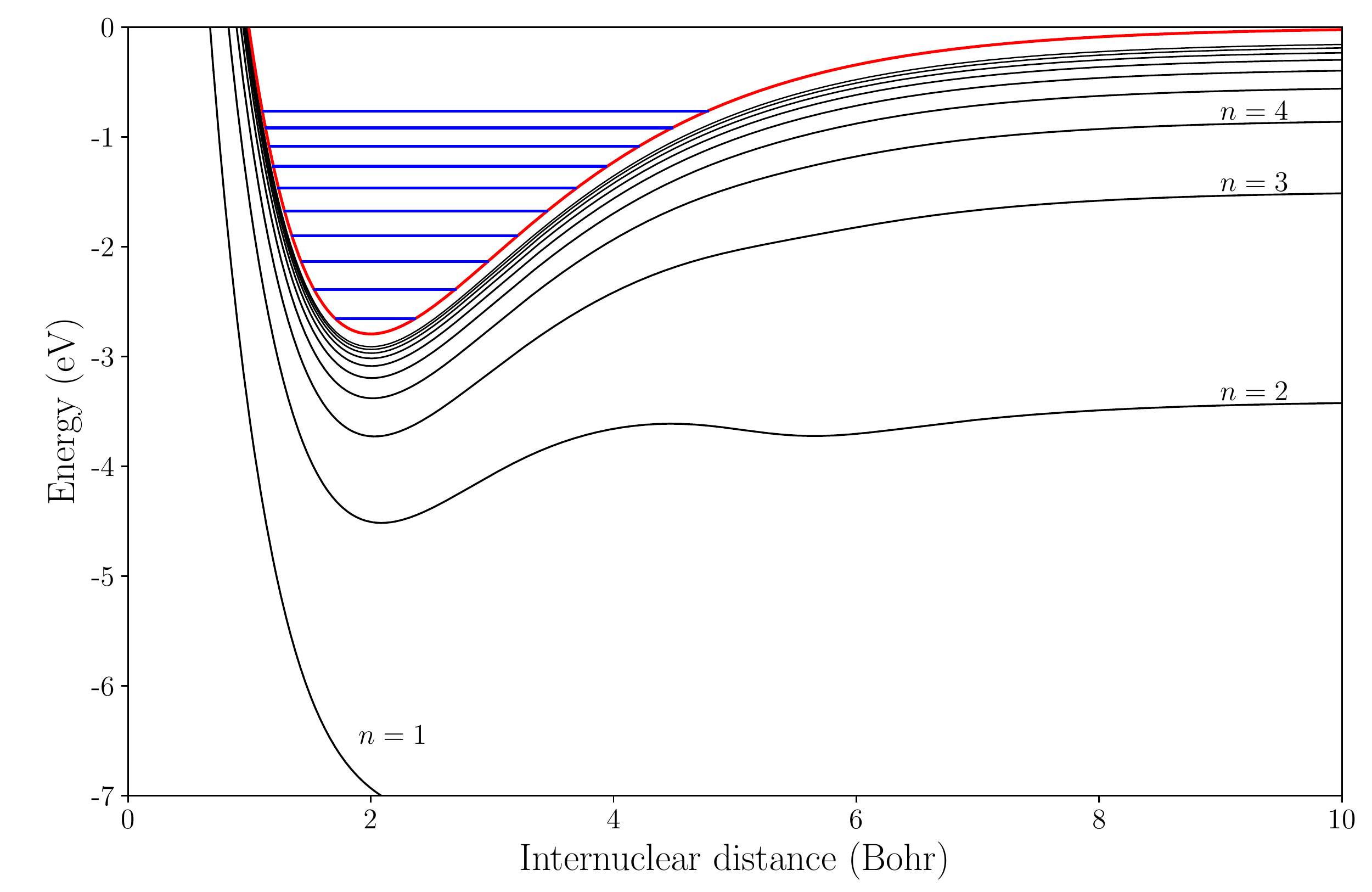}
\caption{\label{fig-PES}
(Color online) Potential energy curves of the H$_2^+$ model.
The curve of $\mathrm{H} + \mathrm{H}^+$ (the potential
$V_0(R)$ in the 2D model) is shown by a red line, while the neutral excited states of $\mathrm{H} +
\mathrm{H}(n)$ are displayed with the black lines. Blue lines display vibrational energy levels of the
cation.
}
\end{figure}

Figs.~\ref{fig-DR} and \ref{fig-PES} show that at zero collision energy, channels with
$n=1,2$ are open. Note that while the model potential (\ref{eq-Ham-pot}) mimics well
the higher-$n$ states of H + H($n p$) system, it becomes less accurate for the lower states.
For example, the model potential supports an uphysical $n=1$, $l=1$ ($p$-wave) for
$R\rightarrow\infty$ with the asymptotic energy of -37.5 eV ($n=1$ curve in 
Fig.~\ref{fig-PES}).
Furthermore, in the examined
collision energy range 0--2 eV another 2 Rydberg channels, $n=3$ and $n=4$, become open
and they are labeled by DR$_3$ and DR$_4$, respectively.

\section{\label{sec-ft}Frame transformation}

We start with the two-dimensional Schr\"{o}dinger equation with the model Hamiltonian
(\ref{eq-Ham-tot})
\begin{equation}
\label{eq-SR_2D}
\left [ -\frac{1}{2} \frac{\partial^2}{\partial r^2}+\frac{l(l+1)}{2r^2}
-\frac{1}{r} -E+H_N(R)+V(R,r) \right ] h(R,r) = 0.
\end{equation}
Following the standard FT approach \cite{Chang_Fano_1972}, this equation can be
solved inside a sphere $r \leq r_0$
that confines the electronic coordinate near the molecule. Within such confinement the
internuclear distance $R$ is a good quantum number and the Born-Oppenheimer approximation holds.
Since we assume that the interaction is characterized by a pure Coulomb potential outside the
sphere ($V(R,r)=0$ for $r \geq r_0$), the Schr\"{o}dinger equation (\ref{eq-SR_2D}) becomes
separable (in the electronic and nuclear coordinates) for $r \geq r_0$.
Consequently, the $j'$-th independent solution of (\ref{eq-SR_2D}) can be
written as a linear combination of the two separable solutions
$\chi_j(R) f^-_j(r)$, $\chi_j(R) f^+_j(r)$ as
\begin{equation}
\label{eq-hl-sphere}
h_{j'}(R,r) \xrightarrow{r \geq r_0} \sum_j \chi_j(R) \left[
f^-_j(r) \delta_{jj'} - f^+_j(r) S_{jj'}\right]\,,
\end{equation}
where the asymptotic wave functions $f^-_j(r)$ and $f^+_j(r)$ are
incoming- and outgoing-wave Coulomb functions,
respectively. The nuclear functions $\chi_j(R)$ are
the eigenstates of the nuclear Hamiltonian $H_N$, and the matrix elements
$S_{j j'}$
are results of the standard frame-transformation integral \cite{Chang_Fano_1972}
\begin{equation}
S_{j j '} = \int dR \chi_j^*(R) e^{2\pi i \mu(R)} \chi_{j'}(R),
\end{equation}
where $\mu(R)$ is quantum defect and $\chi_j(R)$ satisfy simple orthonormality relations
\begin{equation}
\label{eq-orthon}
\int dR \chi_j^*(R) \chi_{j'}(R) = \delta_{jj'}.
\end{equation}

\subsection{Frame transformation into Siegert pseudostates}

In the Appendix \ref{sec-app} we demonstrate, that it is possible to solve the Schr\"{o}dinger
equation (\ref{eq-SR_2D}) via expansion of its $j'$-th independent solution into a
presumably complete subset of $N$ Siegert pseudostates $\phi_j(R)$
\begin{equation}
\label{eq-ft-h}
h_{j'}(R,r)=\sum_{j=1}^N \phi_j(R) g_{j j'}(r).
\end{equation}
This expansion allows us to solve the two-dimensional equation (\ref{eq-SR_2D}) in
the form of a coupled set of $N$ one-dimensional equations (in the coordinate $r$)
\begin{equation}
\label{eq-SRr}
\left [ -\frac{1}{2} \frac{d^2}{d r^2}+\frac{l(l+1)}{2r^2}-\frac{1}{r}-
(E-\epsilon_j)\right ]g_{jj'}(r) +\sum_{m=1}^{N} V_{jm}(r) g_{mj'}(r) = 0,
\end{equation}
where the exact form of the coupling potential $V_{jm}(R)$ can be found in the Appendix.
Note, that the channel thresholds $\epsilon_j$ and the channel-coupling elements
$V_{jm}$ are complex for the nuclear basis formed from Siegert pseudostates.

Our next step concerns the frame transformation of the $S$ matrix into the basis
of the Siegert pseudostates.
The behavior of the interaction coupling matrix $V_{jm}(r)$ at short distances makes
the numerical solution of (\ref{eq-SRr}) very challenging. The nuclear asymptotic
channel functions $\phi_j(R)$ become strongly coupled when the scattered
electronic coordinate $r$ approaches the molecular target, say for $r \leq r_0$.
In this regime the Born-Oppenheimer quantization defined by the fixed $R$
gives a better description than the nuclear channel
functions $\phi_j(R)$ for real molecular applications, as mentioned above.

The remaining derivation follows the concept for the energy-dependent frame transformation
of Gao and Greene \cite{Gao_Greene_PRAR_1990}.
The $j'$-th independent Born-Oppenheimer solution $h_{j'}(R,r)$ of the full Hamiltonian
(\ref{eq-Ham-tot}) inside the sphere can be written at its boundary
$r=r_0$ as a product of the electronic solution at fixed $R$ and the nuclear wave function
$\phi_{j'}(R)$:
\begin{equation}
\label{eq-BO-in}
h_{j'}(R,r_0) = \phi_{j'}(R)\left[ f^-_{j'}(r_0) - f^+_{j'}(r_0) e^{2\pi i \mu(E_{j'},R)} \right]\;,
\end{equation}
where $f^-_{j'}(r)$ and $f^+_{j'}(r)$ are the incoming- and outgoing-wave Coulomb functions
evaluated with the body-frame momentum $k_{j'}$ = $\sqrt{2 E_{j'}}$ = $\sqrt{2(E-\epsilon_{j'})}$.

In the outer region, for $r \geq r_0$, the solution of (\ref{eq-SR_2D})
is expressed in terms of
two separable solutions $\phi_j(R) f^-_j(r)$ and $\phi_j(R) f^+_j(r)$.
Since there is no special guide to match
the two independent solutions of the inner and outer regions (each uses a different quantization
scheme), one needs to construct a general linear combination of the outer solutions.
Therefore, the matching equation has the following form:
\begin{gather}
\label{eq-FT-Gao-0}
\phi_{j'}(R)\left[ f^-_{j'}(r_0) - f^+_{j'}(r_0) e^{2\pi i \mu(E_{j'},R)} \right] =
\sum_m \phi_m(R) \left[ f^-_m(r_0) A_{m j'} - f^+_m(r_0) B_{m j'} \right]\\
\downarrow F_j [\;.\;] \nonumber\\
\label{eq-FT-Gao-1}
 f^-_{j'}(r_0) \delta_{j j'} - f^+_{j'}(r_0) S_{j j'} = f^-_j(r_0) A_{j j'} -
f^+_j(r_0) B_{j j'}\;,
\end{gather}
where
\begin{equation}
\label{eq-Smatrix}
S_{j j'}= \int_0^a dR \phi_j(R) e^{2\pi i \mu(E_{j'},R)} \phi_{j'}(R) +i\phi_j(a)
\left[\left(K_j-i\frac{d}{dR}\right)^{-1}
e^{2\pi i \mu(E_{j'},R)}\phi_{j'}(R)\right]_{R=a}.
\end{equation}
The form of the functional $F_j [\;.\;]$, acting on the variable $R$ on both sides of
Eq.~(\ref{eq-FT-Gao-0}), is presented in the Appendix.
The coefficients $A_{j j'}$ and $B_{j j'}$ can be obtained by using the facts that the
Wronskians of $f^\pm_j$ are independent of $r$,
\begin{eqnarray}
\nonumber
B_{j j'} &=& \frac{-1}{[f^+_j,f^-_j]} \left\{
[f^-_{j'},f^-_{j}] \delta_{j j'} - [f^+_{j'},f^-_{j}] S_{j j'}
\right\} \xrightarrow{\mathrm{energy-independent}\quad f^+,f^-} S_{j j'}
\\
\label{eq-ABcoef}
A_{j j'} &=& \frac{1}{[f^-_j,f^+_j]} \left\{
[f^-_{j'},f^+_{j}] \delta_{j j'} - [f^+_{j'},f^+_{j}] S_{j j'}
\right\} \xrightarrow{\mathrm{energy-independent}\quad f^+,f^-} \delta_{j j'}.
\end{eqnarray}
where $[f,g] = fg'-gf'$ denotes the wronskian of two functions.

Therefore, if we neglect the energy dependence of the Coulomb functions at the point
$r_0$, i.e. $f^\pm_j = f^\pm_{j'}$,
the LAB-frame
$j'$-th independent solution of (\ref{eq-SR_2D}) can be written, for $r \geq r_0$,
in terms of the body-frame quantum defect $\mu(E_{j'},R)$ as
\begin{equation}
\label{eq-h-asyr}
h_{j'}(R,r) = \sum_j \phi_j(R) \left[
f^-_j(r) \delta_{j j'} - f^+_j(r) S_{j j'} \right]\;.
\end{equation}

Finally, in the nuclear asymptotic region, where the quantum defect $\mu(E_{j'},R)$
becomes $R$-independent atomic quantum defect, the Eq.~(\ref{eq-Smatrix}) can
be simplified to
\begin{equation}
\label{eq-Smatrix0}
S_{j j'}= \int_0^a \mathrm{d}R \phi_j(R) e^{2\pi i \mu(E_{j'},R)} \phi_{j'}(R) +
i \frac{\phi_j(a) e^{2\pi i \mu(E_{j'},a)} \phi_{j'}(a)}{K_j + K_{j'}}\;,
\end{equation}
which is the \emph{ad hoc} formula utilized in some of the previous DR studies
\cite{Hamilton_Greene_PRL_2002,Kokoouline_Greene_PRA_2003,
Curik_Greene_PRL_2007,Curik_Greene_MP_2007,Haxton_Greene_NO2_PRA_2012,
Curik_Gianturco_2013,Curik_Greene_JCP_2017}.
Therefore, it is clear that the formula (\ref{eq-Smatrix0}) generates
mathematically correct coefficients in expansion (\ref{eq-h-asyr}) for
$r > r_0$.
\begin{figure}[tbh]
\includegraphics[width=1\textwidth]{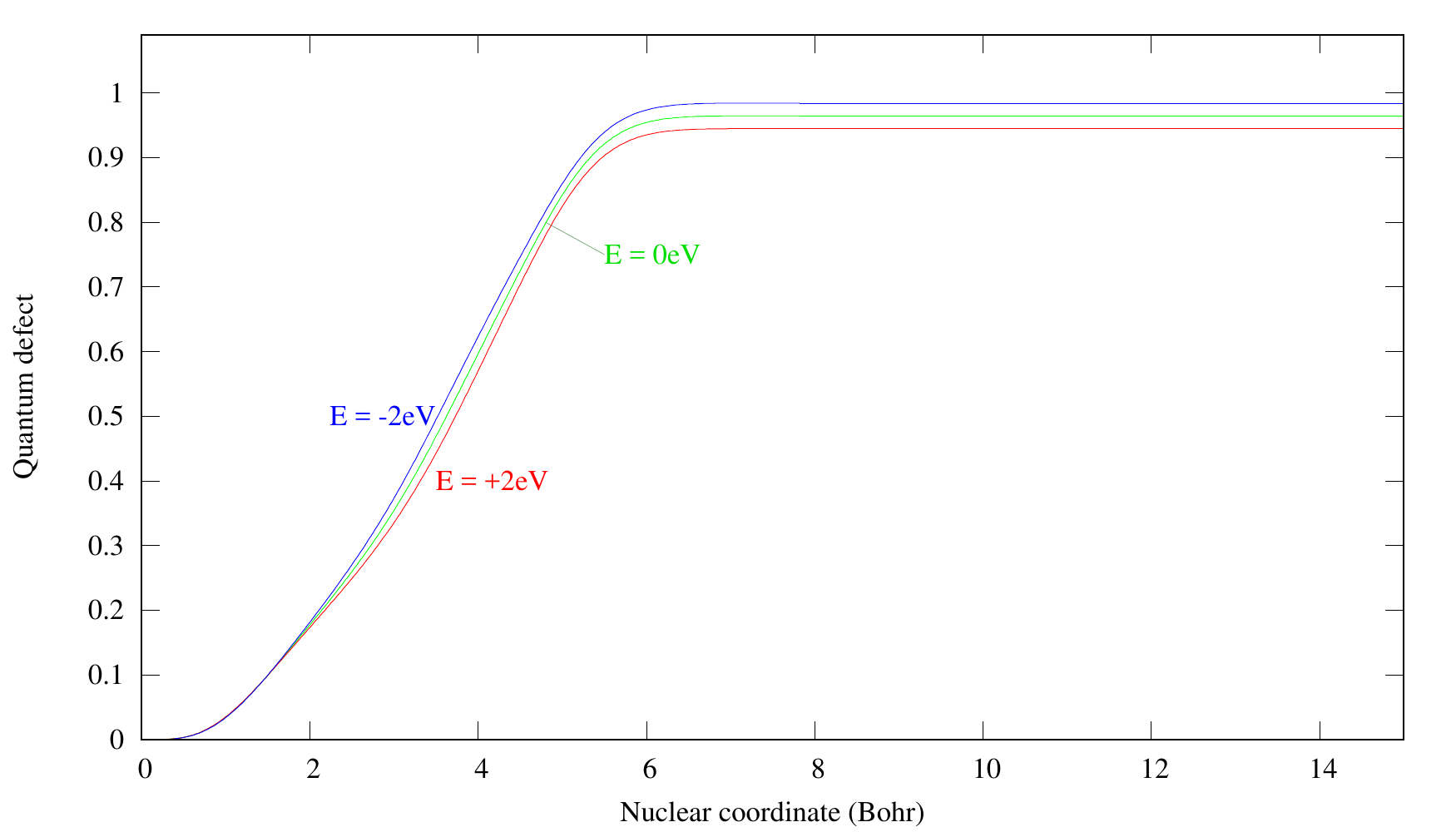}
\caption{\label{fig-QDF}
(Color online) Quantum defect $\mu(E,R)$ of the present model shown for three different
energies, $E = -2$ eV, 0 eV, +2eV.
}
\end{figure}

In order to carry out the vibrational frame transformation (\ref{eq-Smatrix0})
one needs to know the short-range quantum defect $\mu(E,R)$ obtained from
the fixed-nuclei version of the Schr\"{o}dinger equation (\ref{eq-SR_2D}).

Apart from a resonant regime the quantum defect $\mu(E,R)$ is usually a weak
function of energy, owing to the presence of the strong Coulomb field. As
a demonstration, Fig.~\ref{fig-QDF} displays the fixed-nuclei
quantum defect as a function
of nuclear coordinate for three selected energies ($E = -2, 0, 2$ eV).
The quantum defects $\mu(E,R)$ were obtained with a one-dimensional $R$-matrix method
applied to the Hamiltonian $H_e(r) + V_0(R) + V(R,r)$
which is parametrically dependent on $R$.
The negative-energy quantum defects were also
carefully checked against the discrete set of quantum defects that
can be obtained from the discrete set of fixed-nuclei bound-state
energies $E_n(R)$
\begin{equation}
\label{eq-Ryd-eq}
E_n = V_0(R) - \frac{1}{2 \left[n-\mu(E_n, R)\right]^2}\;, n > 1\;.
\end{equation}
The energy dependence of the quantum defect $\mu$ is neglected in the
present study and all the presented results are derived from energy-independent
frame transformation of the zero-energy quantum defect $\mu(0,R)$.

The energy-independent FT presented here consisted of two important steps. In the first
step the energy dependence of the asymptotic Coulomb functions $f^\pm_j$ was neglected
in the Eq.~(\ref{eq-ABcoef}). In the second step we neglected the energy dependence of
the phase shift or quantum defect $\mu(E,R)$ shown in
Fig.~\ref{fig-QDF}. It is important to note that DR is dominated by the quantum
defect values from the Frank-Condon region of the initial vibrational state. In case of
the initial ground vibrational state, centered around 2 Bohrs, we observe very weak
energy dependence of $\mu(E,R)$. However, for higher initial vibrational states that
span to higher values of $R$,
Fig.~\ref{fig-QDF} suggests that the energy dependence of $\mu$ may become
more important.

The final note is made on the third step that should be considered for the energy-independent
FT. Formally, the left side of Eq.~(\ref{eq-FT-Gao-0}) should be multiplied
by a normalization function
$N(E_{j'},R)$ \cite{Gao_Greene_JCP_1989,Gao_Greene_PRAR_1990} because the Born-Oppenheimer
solution inside the electronic sphere $r \leq r_0$ must be normalized independently of $R$.
However, the normalization factor $N(E_{j'},R)$ explodes to infinity in the unphysical
limit in which the energy dependence of $f^\pm_j$ and $\mu(E,R)$ are neglected
simultaneously, as in the present case.  Thus no simple limit for the energy-independent
FT exists here.
Fortunately, it has been shown previously
\cite{Gao_Greene_JCP_1989} that in this case the normalization factor $N$ will drop out
if the vibrational basis is complete. Therefore, since the present derivation was focused
on the introduction of the Siegert pseudostates into the FT theory, we decided to simplify
the equations by omitting the normalization factor from the beginning.

\subsection{Cross section}

It is important to emphasize here that the complex coefficients
(\ref{eq-Smatrix}) in Eq.~(\ref{eq-h-asyr}) are a result of mathematical operations
and calling them $S$-matrix elements in the physical sense may be incorrect. Although
the asymptotic expression (\ref{eq-h-asyr}) for $h_{j'}(R,r)$
takes a familar form, the nuclear states $\phi_j(R)$ are not orthogonal in the conventional sense
(\ref{eq-orthon}). As a consequence, the frame transformed $S$-matrix
elements $S_{j j'}$ do not preserve the original eigenphases -- one of the
properties that was criticized in Ref.~\cite{Haxton_Greene_HeH_2009}.
We believe, that although $S_{j j'}$ probably should not be called $S$-matrix elements,
the asymptotic form (\ref{eq-h-asyr}) is
sufficient to determine the probability flux distribution in different nuclear channels.

As a first step, however, one needs to eliminate the exponentially growing
components of the $h_{j'}(R,r)$ function in (\ref{eq-h-asyr}). This is done by the standard
MQDT technique called "elimination of the closed channels"
\cite{Seaton_RPP_1983,Orange_review}.
This procedure brings a strong energy dependence into the physical $S$ matrix via (apart from a phase
factor)
\begin{equation}
\label{eq-Sphys}
S^{phys}(E) = S_{oo} - S_{oc} \left[ S_{cc} - e^{-2i\beta(E) } \right]^{-1} S_{co},
\end{equation}
where $\beta(E)$ is a diagonal matrix of effective Rydberg quantum numbers with respect to the closed-channel thresholds $\epsilon_j$
\begin{equation}
\beta(E)_{jj'} = \frac{\pi \delta_{jj'}}{\sqrt{2(\epsilon_j-E)}},
\end{equation}
and $S_{oo}, S_{co}, S_{oc}, S_{cc}$ are at present the energy independent sub-matrices of the original energy-independent $S$ matrix
\begin{equation}
        S=\begin{pmatrix}
        S_{oo} & S_{oc}\\
        S_{co} & S_{cc}
\end{pmatrix},
\end{equation}
according to which channels are open and closed at the given energy.
The $r\rightarrow\infty$
asymptote of the total wave-function in the electronic open-channel space,
with an exponential decay in the closed-channel space, can be written as
(apart from a phase factor)
\begin{equation}
\label{eq-Smat-open}
\bar{h}_{j'}(R,r) \xrightarrow{r\rightarrow\infty} \frac{1}{2\pi k_{j'}} \left[
\phi_{j'}(R) e^{-i k_{j'} r}  - \sum_{j=1}^{N_o} \phi_j(R) e^{i k_j r} S^{phys}_{jj'}
+ \sum_{j=N_o+1}^N \phi_j(R) e^{-k_j r} Z_{jj'}
\right],
\end{equation}
where $N_o$ is the number of open channels. The closed-channel coefficients $Z_{j j'}$
can be found in the literature \cite{Orange_review}.

Owing to the complex nature of the Siegert pseudostates the physical $S$ matrix
$S^{phys}$ becomes subunitary. The subunitarity is predominantly a result of
the channel elimination
procedure that combines the channel solutions in such a way that only exponentially
decreasing parts of the electronic wave function survive in the close nuclear
channels represented by the outgoing Siegert pseudostates with a finite lifetime.
While the third term on the r.h.s. of (\ref{eq-Smat-open}) clearly represents a portion of the
total wave function in which the electron is described by a combination of the exponentially
decaying functions with the nuclear components having the outgoing-wave boundary conditions,
at present we are not able to recast this term into Eq.~(\ref{eq-wfn-2d-asy})
in which the electronic energies are discrete (atomic Rydberg states) and
the nuclear energies are continuous (nuclear kinetic energy release).

Hamilton and Greene \cite{Hamilton_Greene_PRL_2002} postulated that all
of the probability flux lost from the open channels can be associated
with a trapped Rydberg electron in closed channels that represent
dissociative states. Thus the subunitarity of $S^{phys}$ is caused by missing dissociation
probability following electron impact in the incident channel $j'$
\begin{equation}
\label{eq-dunit}
\sigma_{j'}(E) = \frac{\pi}{2(E-\epsilon_{j'})}
\left[ 1 - \sum_j S^{phys}_{jj'}(E) S^{phys\,\dag}_{j'j}(E) \right].
\end{equation}
This cross section does not differentiate between outgoing Rydberg channels and thus it represents the DR cross section summed over all the final Rydberg channels (electronic atomic states).

The goal of the following section is to provide a numerical evidence for the above described
physically sound, yet intuitive approach to compute dissociative recombination
cross sections.

\section{\label{sec-res}Results and discussion}

\begin{figure}[!t]
\centering
\includegraphics[width=1\textwidth]{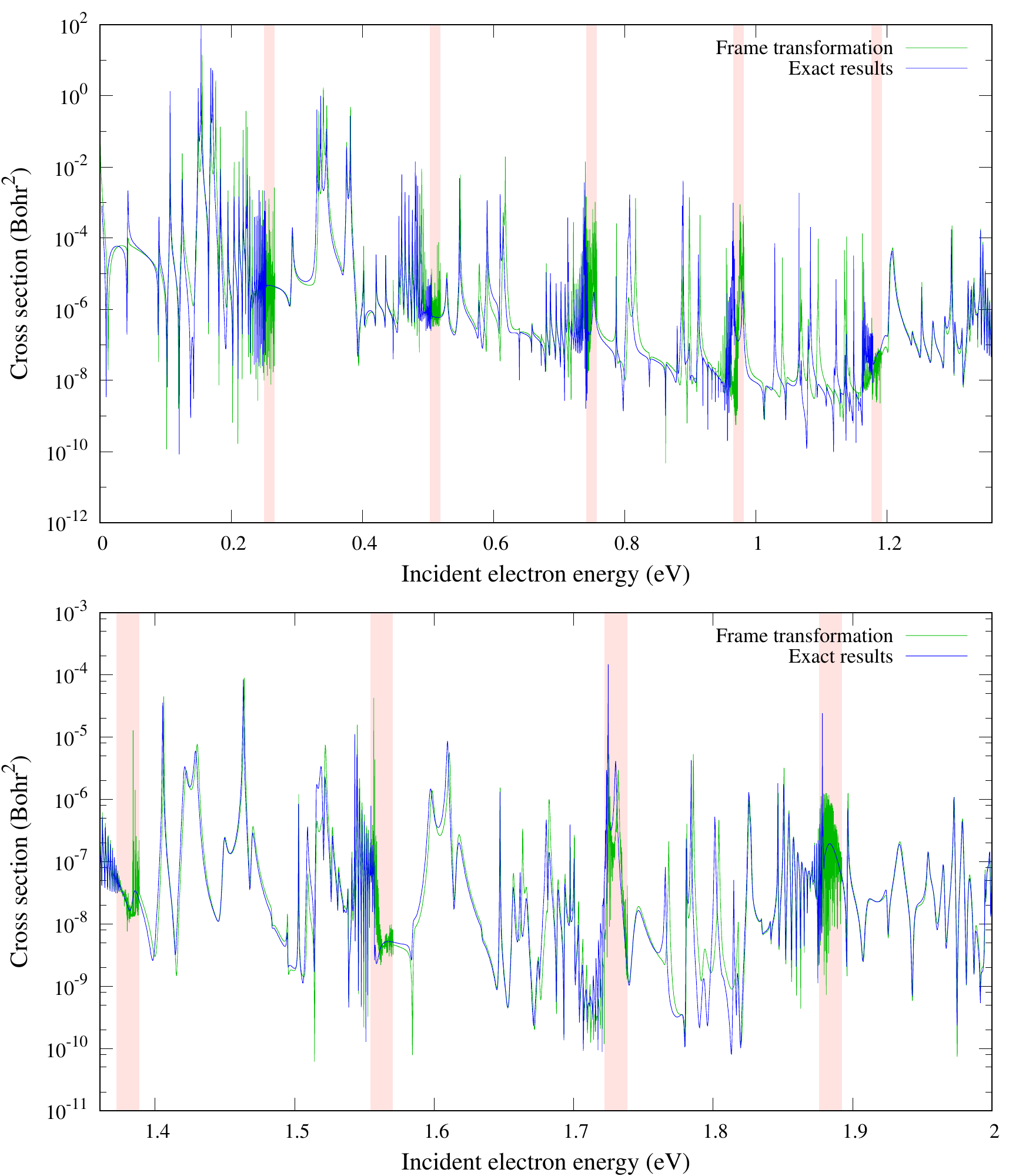}
\caption{\label{fig-com-full}
(Color online) Comparison of the calculated cross sections for
the collision energies from 0 eV to 2 eV.
The total DR cross section obtained with the frame transformation approach is shown with the
green curve. The exact results represent a sum of the DR cross sections over the open electronic
(Rydberg) channels computed with the
2D model. Note that the exact results are not
fully converged in the pink areas right below vibrational excitation
thresholds, see Sec.~\ref{sec-num-imp} for explanation.
}
\end{figure}

To test the theory presented in the Section~\ref{sec-ft} we compare the frame
transformation (FT) results for the two-dimensional model introduced in Section~\ref{sec-exmod} with
exact ones obtained from a direct solution of this model. The results are compared for electron
collision energies in the interval from 0 to 2 eV. This energy range spans over
nine vibrational thresholds and contains two Rydberg thresholds for $n = 3$ and
$n = 4$ states (see Fig.~\ref{fig-PES}) of the neutral H($n$) atom after the
model dissociation
\begin{equation}
\mathrm{H}_2^+(^2\Sigma_g^+) + e^-(l=1,m=0) \rightarrow
\mathrm{H}(1s) + \mathrm{H}(np) \;.
\end{equation}

\begin{figure}[th]
\centering
\includegraphics[width=0.85\textwidth]{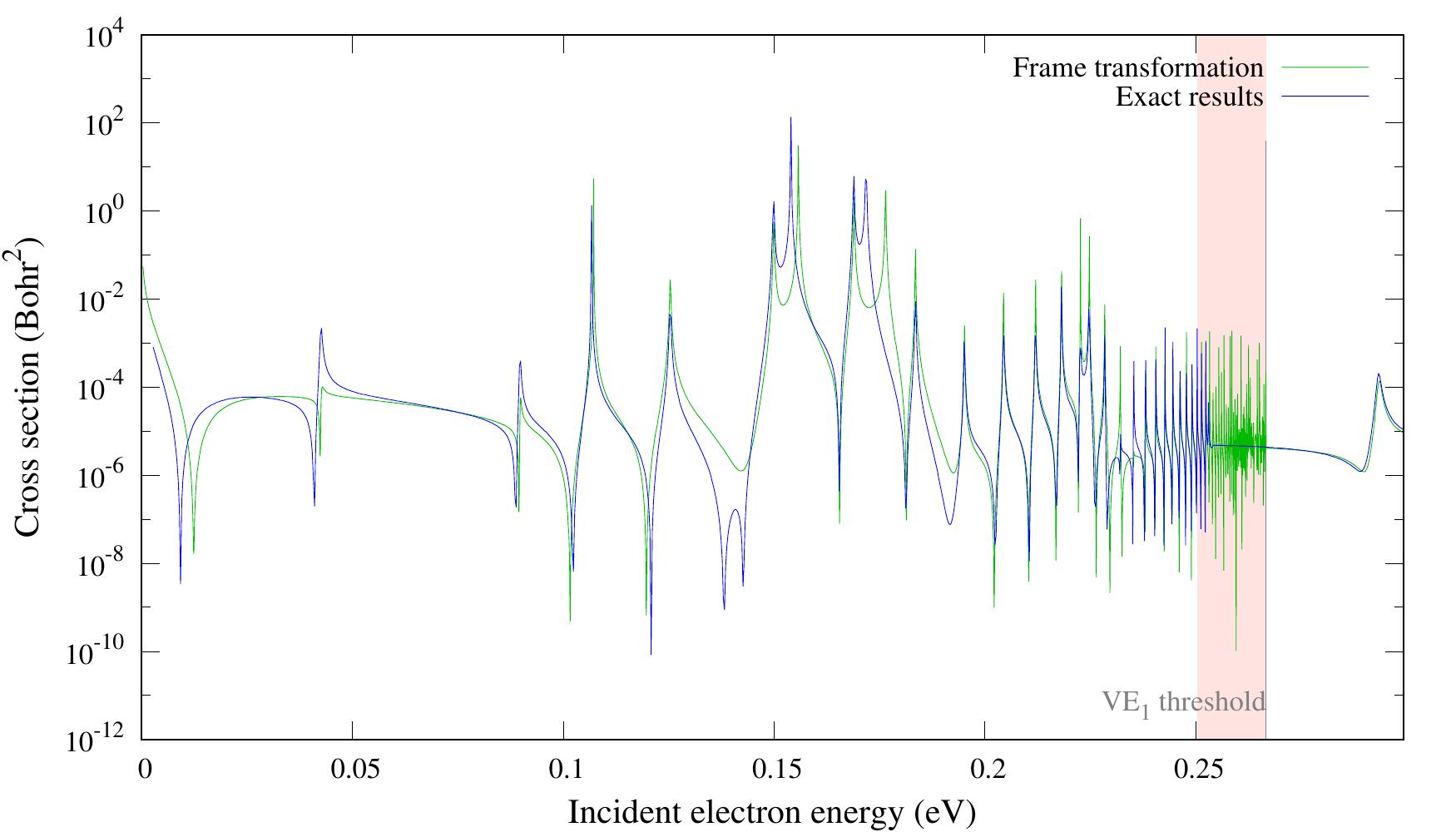}
\caption{\label{fig-zoom1}
(Color online) Comparison of the calculated cross sections for the collision energies
up to the first
vibrational threshold. The colors used are the same as in Fig.~\ref{fig-com-full}.
}
\end{figure}
\begin{figure}[th]
\centering
\includegraphics[width=0.85\textwidth]{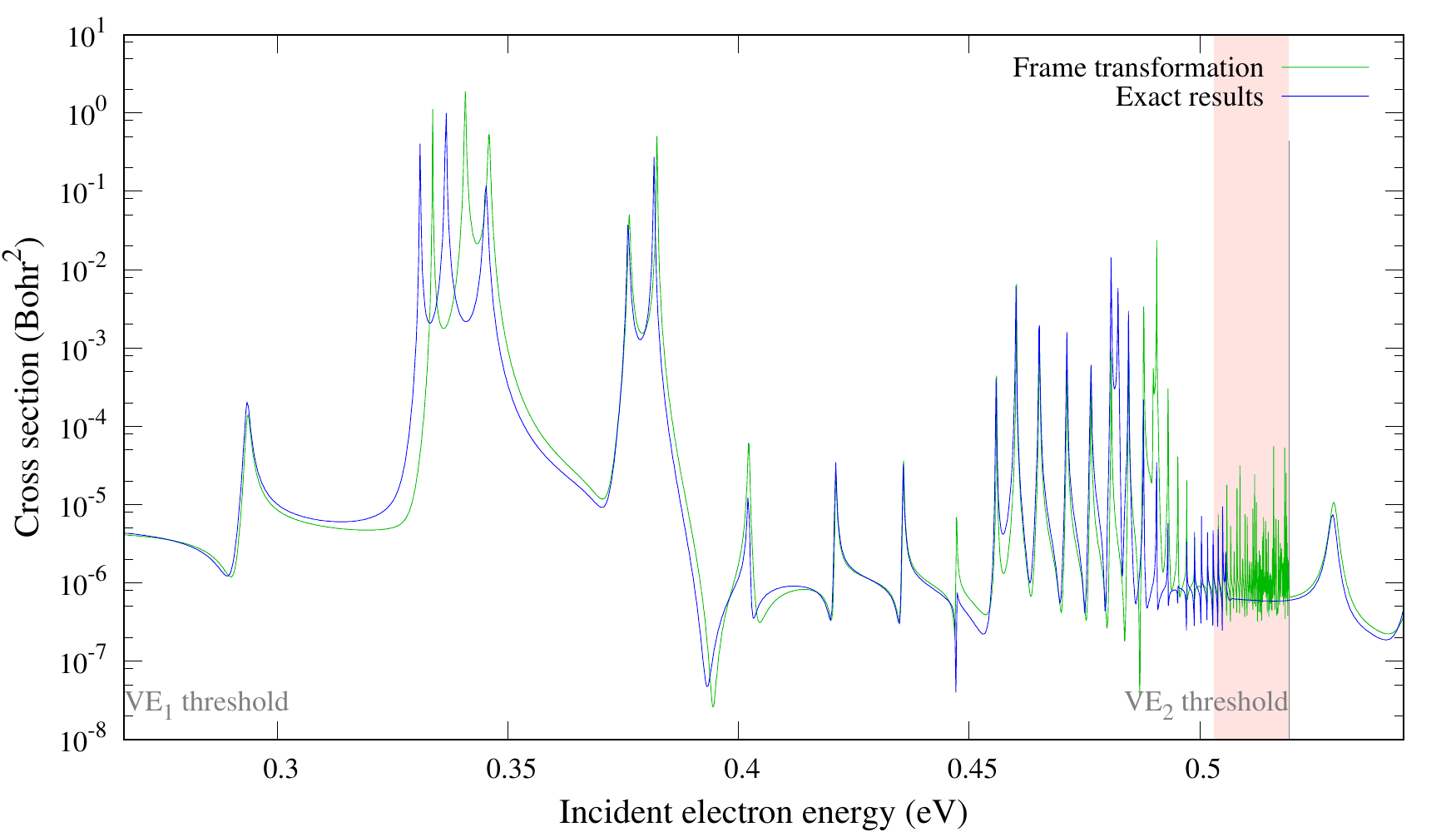}
\caption{\label{fig-zoom2}
(Color online) Comparison of the calculated cross sections for the collision energies between
the first and second vibrational thresholds.
The colors used are the same as in Fig.~\ref{fig-com-full}.
}
\end{figure}

At first glance at Fig.~\ref{fig-com-full}, there is a good correspondence between the FT calculated
total DR cross section and the exact results for the 2D model. As described in Section~\ref{sec-exmod}
the pink rectangles highlight the areas below the vibrational thresholds where the FEM-DVR-ECS results
for the 2D model are not converged because the electron is trapped in high-$n$ Rydberg states which do
not fit into the chosen electronic grid. Thus to assess validity of the FT approach one should compare
the results only outside of these pink rectangles where the data are converged with respect to all the
parameters present in the numerical implementations, i.e. with respect to nuclear
and electronic grid sizes and corresponding grid densities. In case of the FT method with Siegert
pseusostates, these parameters contain the nuclear grid size $a$ and the number
of Siegert pseudostates included in the frame transformation, i.e. the size of the
$S$ matrix (\ref{eq-Smatrix0}).

\begin{figure}[th]
\centering
\includegraphics[width=0.85\textwidth]{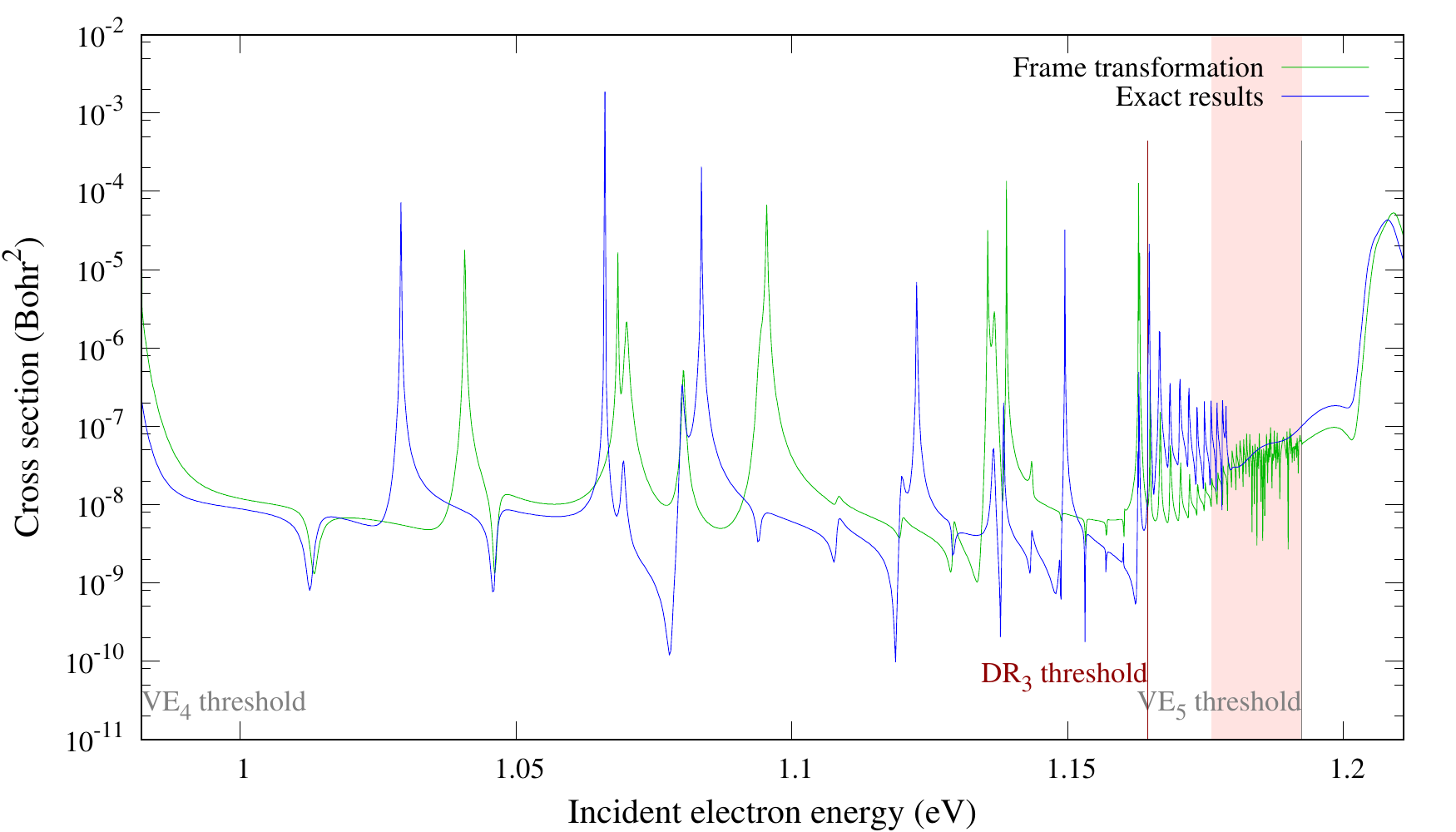}
\caption{\label{fig-zoom5}
(Color online) Comparison of the calculated cross sections for the collision energies between
the fourth and fifth vibrational thresholds.
The colors used are the same as in Fig.~\ref{fig-com-full}.
}
\end{figure}

For a more detailed analysis we present several zoomed graphs. In Figs.~\ref{fig-zoom1} and
\ref{fig-zoom2} we show that agreement between the FT and exact results is
very good. The FT approach together with the hypothesis centered around
Eq.~(\ref{eq-dunit}) seems to count the DR flux correctly and most of the resonant
features are accounted for.

\begin{figure}[th]
\centering
\includegraphics[width=0.85\textwidth]{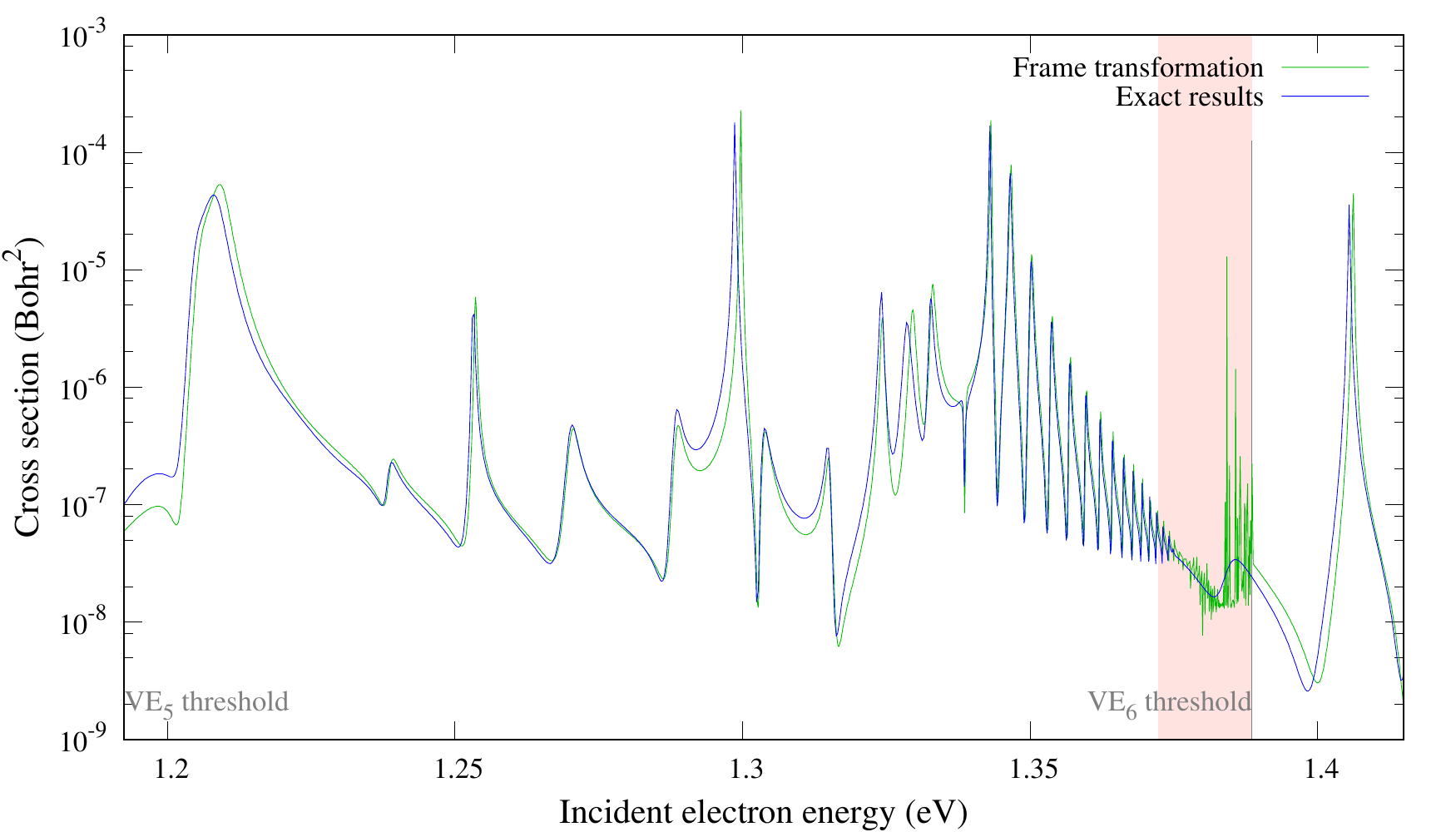}
\caption{\label{fig-zoom6}
(Color online) Comparison of the calculated cross sections for the collision energies between
the fifth and sixth vibrational thresholds.
The colors used are the same as in Fig.~\ref{fig-com-full}.
}
\end{figure}

\begin{figure}[th]
\centering
\includegraphics[width=0.85\textwidth]{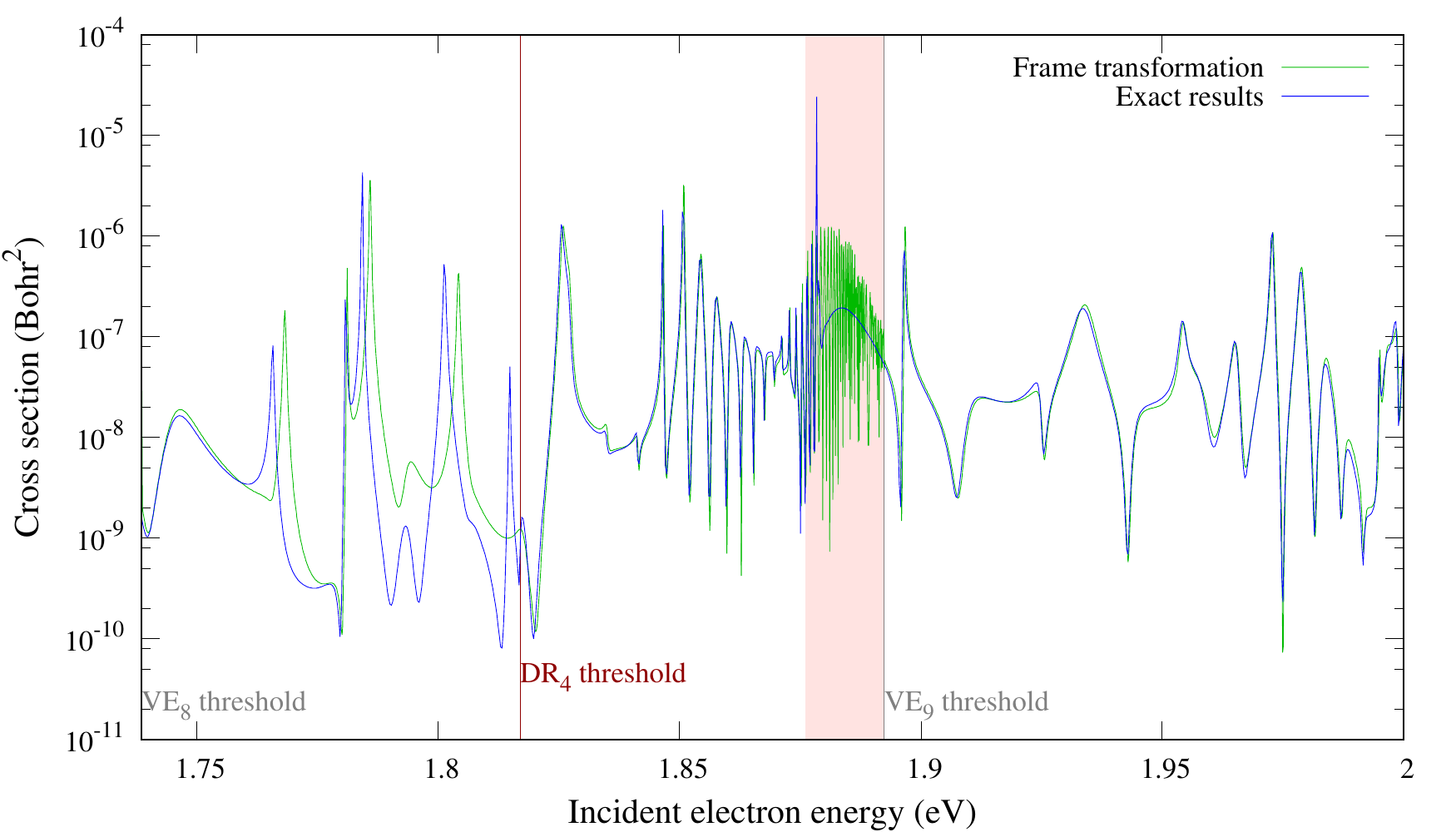}
\caption{\label{fig-zoom9}
(Color online) Comparison of the calculated cross sections for
the collision energies between the eighths and ninth
vibrational thresholds. The colors used are the same as in Fig.~\ref{fig-com-full}.
}
\end{figure}
\begin{figure}[htb]
\centering
\includegraphics[width=0.85\textwidth]{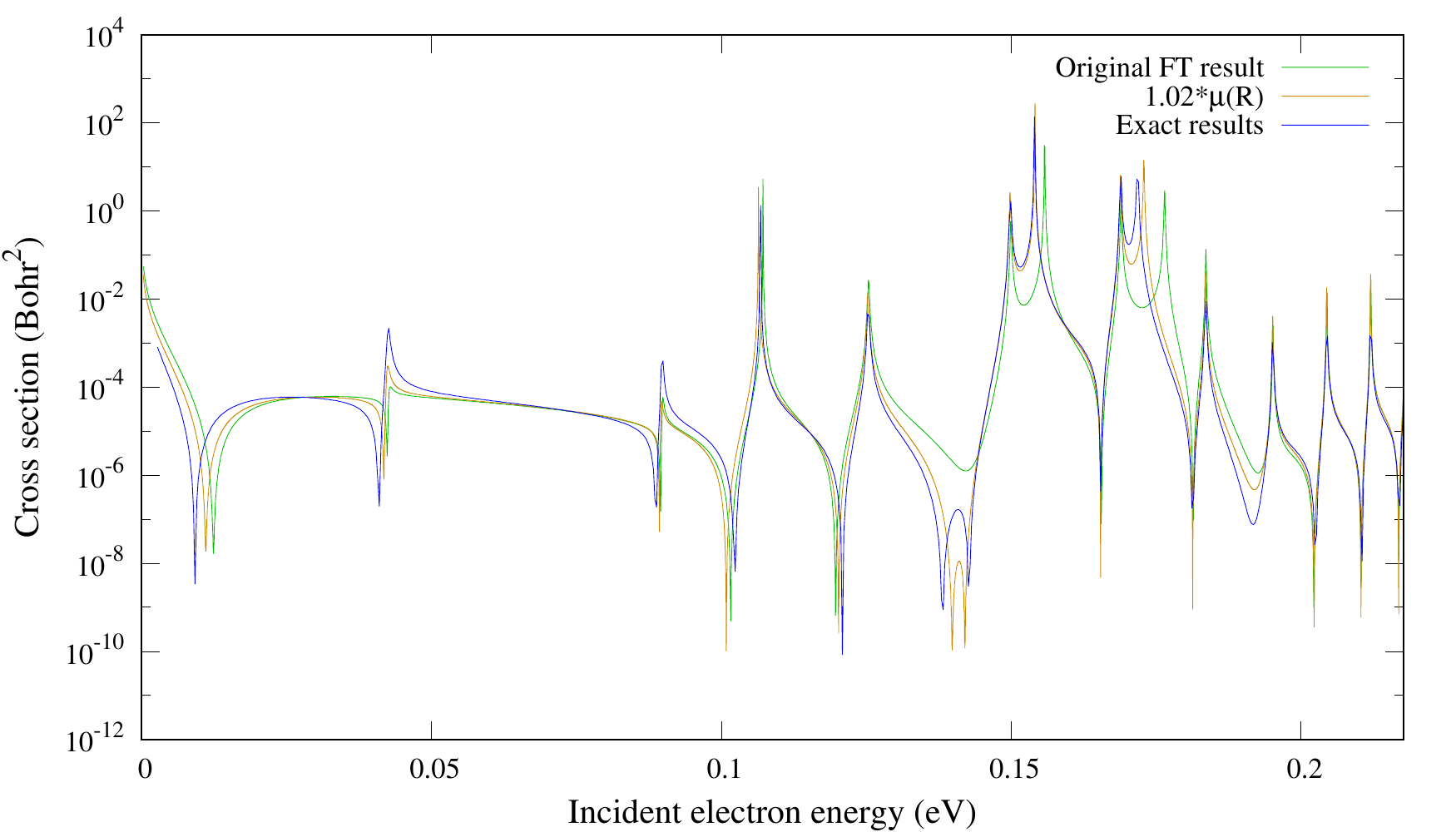}
\caption{\label{fig-defect}
(Color online) Comparison of the calculated cross sections for the
collision energies from 0 to 200 meV.
The total DR cross section obtained with the FT approach is shown with the
green curve. The corresponding exact results are displayed by the blue curve.
The cross section for artificially increased (by 2\%) quantum defect is shown by the
orange curve.
}
\end{figure}

\begin{figure}[!ht]
\centering
\includegraphics[width=0.85\textwidth]{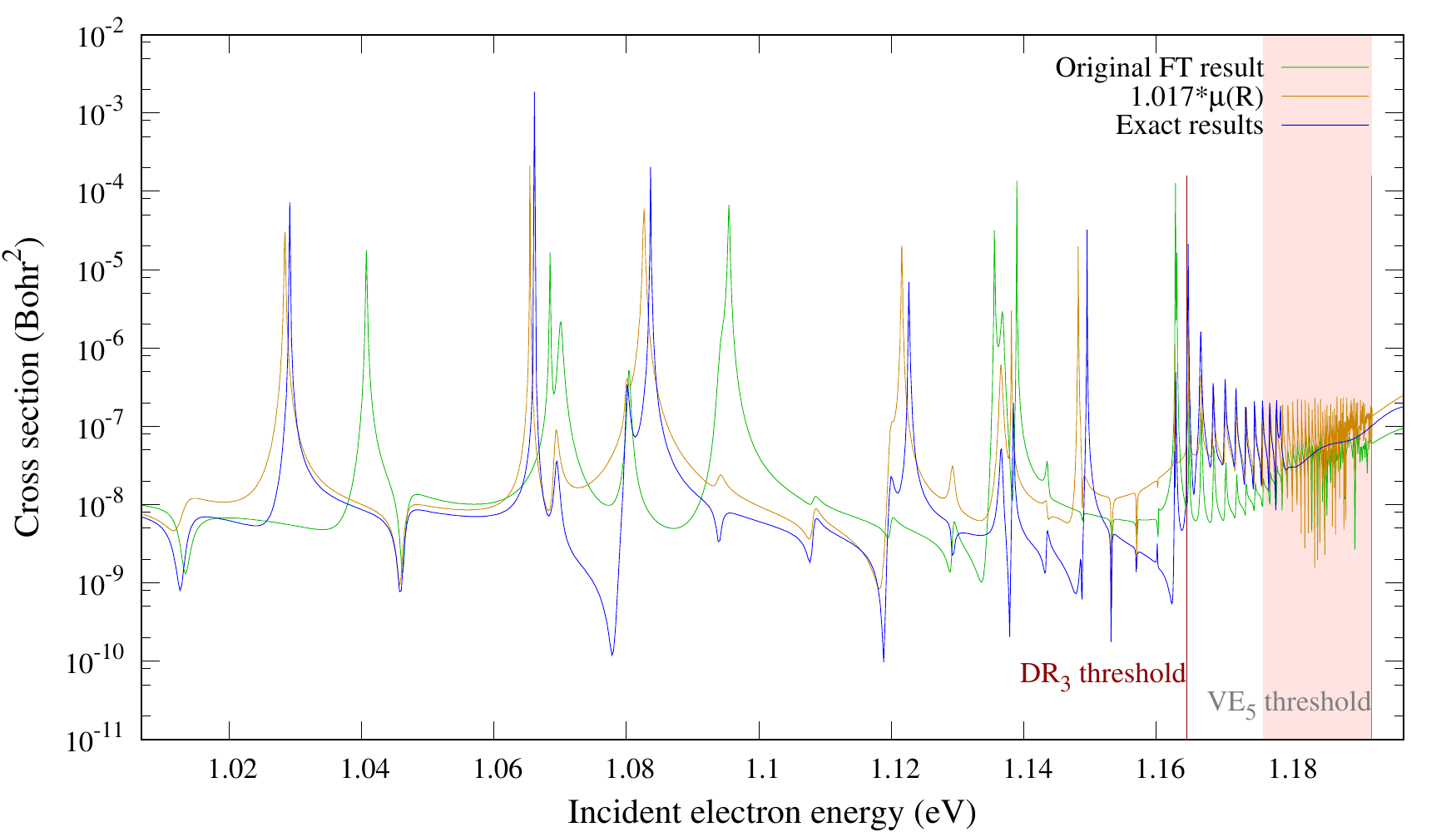}
\caption{\label{fig-defect2}
(Color online) Comparison of the calculated cross sections for the collision energies below the
$n = 3$ Rydberg threshold.
The total DR cross section obtained with the FT approach is shown with the
green curve. The corresponding exact results are
displayed by blue curve.
The cross section for artificially increased (by 1.7\%) quantum defect is shown by the
orange curve.
}
\end{figure}

The comparison becomes worse for the collision energies
just under the opening of the first ($n = 3$) Rydberg threshold. The corresponding
energy window between the fourth and fifth vibrational thresholds is
displayed in Fig.~\ref{fig-zoom5}. Upon opening the $n = 3$ Rydberg threshold, the
exact DR cross section rises sharply and the highest
open Rydberg channel becomes dominant. The corresponding
FT cross section does not distinguish between different electronic channels as it accounts
only for a summed probability via the formula (\ref{eq-dunit}).
Although the FT data do not raise so sharply above the $n = 3$ threshold, they increase for
slightly higher energies giving good agreement with the exact results again.
Corresponding comparison continues on Fig.~\ref{fig-zoom6}.

Similar behavior is also observed in the vicinity of the $n = 4$ threshold, displayed in
Fig.~\ref{fig-zoom9}. Below this threshold a visible disagreement between the FT
and exact results can be seen, while above the threshold the FT approach works very well again.

In order to assess the origin of the small discrepancies between the
employed method we carried a simple computational experiment in which
we recalculated the cross section in the first energy window (0--200 meV)
with a quantum defect $\mu(R)$ that was artificially increased by 2\%.
The size and sign of the increase corresponds to typical energy
dependence of the quantum defect we observe in the present interval
of collision energies as can be seen in Fig.~\ref{fig-QDF}.
Results of this computational experiment are shown in Fig.~\ref{fig-defect}.
Comparison with the exact results shows that the cross section for the artificially
increased quantum defect becomes closer to the exact values.
Even the double resonant structure at 140 meV (blue line in Fig.~\ref{fig-defect})
is reconstructed by this computational experiment.
It is clear that the small deviations of the FT approach from the exact
results can be easily explained by the energy dependence of the quantum defect,
an effect not included in the present study.

Furthermore, the following Fig.~\ref{fig-defect2} demonstrates that even
larger discrepancies between two approaches, seen under the $n = 3$
Rydberg threshold, can also be explained by energy dependence of the quantum
defect. In Fig.~\ref{fig-defect2} we show results of a similar experiment
in which we again artificially increase the quantum defect to simulate
its energy dependence. In this energy range, under the opening of the
$n = 3$ channel, the DR cross section appears to be very sensitive to
accuracy of the quantum defect. The artificial results are much closer
to the exact values and even the DR$_3$ threshold behavior of the
FT approach becomes almost exact.

\section{\label{sec-con}Conclusions}

The aim of the present study is twofold. Firstly, it is to present
modification of the numerically solvable two-dimensional model
of electron-molecule collisions for application to
dissociative recombination problems. Secondly, it attempts to justify two
theoretical steps in the frame transformation approach that uses a basis of
Siegert pseudostates.

The 2D model simulates collisions of
electrons with H$_2^+$ ions leading to the computationally most challenging
channel, the dissociative recombination. Apart from narrow energy windows
right below each vibrational threshold, we were able to obtain accurate
and converged results in the collision energy range from 0 eV to 2 eV.
The DR process in these narrow energy windows is governed by high-$n$
Rydberg states that do not fit into our limited electronic grid size.
Therefore, while in theory these unconverged energy windows can be made
arbitrarily small (in expense of CPU time), they cannot be completely removed.
Nonetheless, the 2D model presented in this work, allowed
the first DR study that does not take into account any kind of approximation
in electron-nuclear interactions. As such, it serves here (and may serve
in the future) as a useful tool for benchmarking various frame
transformation methods developed in the past or possible approaches
designed in the future.

Frame transformation in combination with Siegert pseudostates was previously applied
to a number of molecular targets (for the detailed list see the
Sec.~\ref{sec-intro}). While some of the theory's footings were intuitive,
the studies provided high-quality DR cross sections that reproduced experimental
data well.
With respect to the frame transformation theory we have demonstrated
that under reasonable physical
assumptions the frame-transformed $S$ matrix (\ref{eq-Smat-FT}) provides
coefficients that combine nuclear channels represented by the complex Siegert
pseudostates with $S$-matrix asymptotes in the electronic coordinate. Such channels
are not orthogonal in the conventional sense and therefore the resulting
DR probabilities were judged by a numerical comparison against the exact results
of the 2D model.

The dissociative recombination cross sections resulting from the two
approaches are found to be in very good agreement for all the collision
energies from 0 to 2 eV, except two energy windows that are placed just
below the Rydberg thresholds defining the openings of channels with higher
electronic state $n$ of the final hydrogen atom. Via a simple computational
experiment we have shown that the larger discrepancies in these problematic
energy windows may be easily explained by an energy dependence of the
quantum defect, a feature that is neglected in the present study. Moreover,
the small differences between the FT and exact results over all the
studied energy range from 0 to 2 eV can also be explained by this neglected
energy dependence.

The second possible origin of the discrepancies may lie
in the inaccuracy of the Born-Openheimer approximation that is exploited
(at the short range) by the present FT theory while the 2D model
is free of such approximations. Hence, the following step in this project
would be to consider the weak energy dependence of the wave functions
merged at the FT boundaries and attempt to obtaid the DR rates by so-called
energy-dependent frame transformation.
However, an implementation of the energy-dependent frame transformation
for the dissociative recombination processes appears to be a non-trivial problem and
we plan to make it subject of a future, separate study.

%
\appendix
\section{\label{sec-app}Expansion into a nuclear basis represented by Siegert pseudostates}

We aim to solve the two-dimensional Schr\"{o}dinger equation
\begin{equation}
\label{eq-ap-SR_2D}
\left [ -\frac{1}{2} \frac{\partial^2}{\partial r^2}
+\frac{l(l+1)}{2r^2}-\frac{1}{r}-E+H_N(R)+V(R,r) \right ] h(R,r) = 0,
\end{equation}
where $H_N(R)$ is the nuclear Hamiltonian
\begin{equation}
\label{eq-nuc1}
H_N(R)=-\frac{1}{2 M} \frac{\partial^2}{\partial R^2}+V_0(R)
\end{equation}
and the coupling potential $V(R,r)$ is defined by Eq.~(\ref{eq-Ham-pot}).
Let us now assume that we have already solved the one-dimensional nuclear problem
\begin{equation}
\label{eq-ap-SPS_0}
H_N(R)\phi_j(R)=\epsilon_j \phi_j(R),
\end{equation}
with the boundary conditions
\begin{eqnarray}
\nonumber
\phi_j(0) &=& 0,\\
\left. \left ( \frac{d}{dR} - iK_j \right )
\phi_j(R) \right|_{R=a} &=& 0,
\label{eq-ap-SPS_1}
\end{eqnarray}
where $\epsilon_j=\frac{K_j^2}{2 M}$ and $a$ is some finite distance.
These boundary conditions define a basis of Siegert pseudostates
\cite{Tolst_sieg_1998}. Their orthogonality relations are
\begin{equation}
\label{eq-ap-OG_1}
\int_0^a \phi_j(R) \phi_{j'}(R) dR +i\frac{\phi_j(a) \phi_{j'}(a)}{K_j+K_{j'}}
= \delta_{jj'}.
\end{equation}

The orthogonality relations are valid for all the $2 \times N_b$ pseudostates,
that are provided by $N_b$ basis set elements. It is clear that the
full set of $2 \times N_b$ Siegert pseudostates is overcomplete, in fact it
spans the Hilbert space of the original basis set exactly twice
\cite{Tolst_sieg_1998}.
Out of this overcomplete set we select a subset of $N$ Siegert pseudostates that
contains all the bound states plus the outgoing-wave continuum states. This subset
can be made complete to a sufficient numerical accuracy.

Assuming we have selected the complete subset
of $N$ Siegert pseudostates
$\phi_j(R)$ satisfying Eqs.~(\ref{eq-ap-SPS_0}) and (\ref{eq-ap-SPS_1})
the $j'$-th independent solution $h_{j'}(R,r)$ of (\ref{eq-ap-SR_2D}) can be
expanded as
\begin{equation}
\label{eq-ap-h}
h_{j'}(R,r)=\sum_{j=1}^N \phi_j(R) g_{j j'}(r).
\end{equation}
Inversion of this equation and determination of the expansion coefficients
$g_{j j'}(r)$ is not straightforward due to the non-trivial orthogonality
relations (\ref{eq-ap-OG_1}). One needs to construct a well behaved
one-index linear functional $F_j[\;.\;]$, acting on the space of functions
$f(R)$, that satisfies
\begin{equation}
\label{eq-app-Fdelta}
F_j[\phi_{j'}]=\delta_{j j'} \hbox{ for all } j,j'.
\end{equation}
Basicaly, the functional $F_j[\;.\;]$ applied to $\phi_{j'}$ needs to replicate the left side
of (\ref{eq-ap-OG_1}).

One of the ways to define the functional $F_j[\;.\;]$ is
\begin{equation}
F_j[f(R)]= \int_0^a dR \phi_j(R) f(R) +i\phi_j(a)
\left[\left(K_j-i\frac{d}{dR}\right)^{-1} f(R)\right]_{R=a},
\end{equation}
which indeed simulates (\ref{eq-ap-OG_1}) when applied to $\phi_{j'}(R)$.
The inverted operator on the r.h.s. of the equation above needs to be interpreted
as a complex function of the nuclear operator $D = -id/dR$. Since the
operator $D$ is non-hermitian on the class of functions that are non-zero at $R=a$,
one needs to resort to its definition by a Taylor expansion
\begin{equation}
\label{eq-app-tay1}
\left(K_j+D\right)^{-1}=
\frac{1}{K_j}-\frac{1}{K_j^2}D+\frac{1}{K_j^3}D^2-\ldots
\end{equation}
Even if we skip the discussion of a convergence of this expansion we still need the
Siegert pseudostates $\phi_{j'}(R)$ to satisfy
\begin{equation}
\label{eq-app-Dl}
\left. D^p \phi_{j'}(R)\right|_{R=a} = K^p_{j'} \phi_{j'}(a)\;.
\end{equation}
This is trivially satisfied for $p=1$ by the boundary condition (\ref{eq-ap-SPS_1}).
The second derivative can be obtained from the Schr\"{o}dinger equation
(\ref{eq-ap-SPS_0}) as
\begin{equation}
\label{eq-app-D2}
\left. D^2 \phi_{j'}(R)\right|_{R=a} = K^2_{j'}\phi_{j'}(a) - 2 M V_0(a)\phi_{j'}(a)\;,
\end{equation}
where the potential term $-2 M V_0(a)$ can be made arbitrarily small by increasing
the boundary $a$. All the higher
order derivatives in (\ref{eq-app-tay1}) can be generated by a combination of the boundary
condition (\ref{eq-ap-SPS_1}) and multiple applications of (\ref{eq-app-D2}).
The deviations from the sought property (\ref{eq-app-Dl}) will be proportional
to the surface value of the potential $V_0(a)$ and its higher-order derivatives at the
boundary $a$. In
the present case of the Morse potential $V_0$ we were able to obtain the expected
properties (\ref{eq-app-Dl}) and (\ref{eq-app-Fdelta}) within a very good numerical
accuracy for $a \geq 20$ Bohrs. Realistic target cations involve asymptotically the
induced dipole interaction behaving as $\sim -1/R^4$. However, such asymptote
leads to decreasing higher-order derivatives at the boundary and the sought
property (\ref{eq-app-Dl}) can be easily satisfied.

Application of $F_j[\;.\;]$ then simulates
the typical projection $\int \mathrm{d}R \phi^*_j(R)\ldots$ used for the conventional
orthonormality relations. Consequently, the $F_j[\;.\;]$ allows to invert
Eq.~(\ref{eq-ap-h}) giving
\begin{equation}
g_{j j'}(r)=F_j[h_{j'}(R,r)]=\int_0^a dR \phi_j(R) h_{j'}(R,r) +i\phi_j(a)
\left[\left(K_j-i\frac{d}{dR}\right)^{-1} h_{j'}(R,r)\right]_{R=a}.
\end{equation}
Furthermore, application of the functional onto the both sides of the two-dimensional
Schr\"{o}dinger equation (\ref{eq-ap-SR_2D}) leads to a set of coupled one-dimensional
equations via
\begin{gather}
\left [ -\frac{1}{2} \frac{d^2}{d r^2}+
\frac{l(l+1)}{2r^2}-\frac{1}{r}- E +H_N(R)+V(R,r) \right ] h_{j'}(R,r) = 0, \nonumber \\
\downarrow \hbox{Eq.~(\ref{eq-ap-h})}
\nonumber\\
\sum_m \left [ -\frac{1}{2} \frac{d^2}{d r^2}+
\frac{l(l+1)}{2r^2}-\frac{1}{r}- E +\epsilon_m\right] \phi_m(R) g_{m j'}(r) +
\sum_m V(R,r)\phi_m(R) g_{m j'}(r) = 0, \nonumber\\
\downarrow F_j[\;.\;]\nonumber\\
\label{eq-ap-gii}
\left [ -\frac{1}{2} \frac{d^2}{d r^2}+
\frac{l(l+1)}{2r^2}-\frac{1}{r}-( E -\epsilon_j)\right ]g_{j j'}(r) +
\sum_m V_{jm}(r) g_{m j'}(r) = 0,
\end{gather}
where
\begin{equation}
V_{jm}(r)= \int_0^a dR \phi_j(R) V(R,r)\phi_m(R) +i\phi_j(a)
\left[\left(K_j-i\frac{d}{dR}\right)^{-1}V(R,r)\phi_m(R)\right]_{R=a}.
\label{eq-ap-coupling}
\end{equation}
The coupling interaction elements $V_{jm}$ contain an additional surface term which again
can be made arbitrarily small in practical applications, by increasing the radius $a$.

Radial solutions $g_{j j'}(r)$ of the coupled system (\ref{eq-ap-gii}) form the full
two-dimensional solution $h_{j'}(R,r)$ via the expansion (\ref{eq-ap-h}) as long as this
expansion is complete. This procedure is called vibrational
close-coupling expansion in the literature \cite{Morrison_greenbook1} and its main
objective is the numerical solution of the coupled set of equations (\ref{eq-ap-gii}) from
$r = 0$ to $r = r_0$, beyond which $V(R,r) = 0$.

To summarize, we have just demonstrated that the vibrational close-coupling procedure can also employ
the non-orthogonal system of complex Siegert pseudostates $\phi_j(R)$ and the two-dimensional solution
$h_{j'}(R,r)$ can be reconstructed from its complete subset. The channel-coupling potential elements
$V_{jm}(r)$ (\ref{eq-ap-coupling}) have a typical $C$-norm form -- no conjugation on the bra-element
with an additional surface term that can be made arbitrarily small.

\begin{acknowledgments}
R\v{C} conducted this work within the COST Action CM1301
(CE\-LI\-NA) and the support of the Czech Ministry of Education (Grant No. LD14088) is
acknowledged. MV and KH acknowledge financial support from the Grant Agency of Czech Republic under
contract number GACR 16-17230S. The contributions of CHG were supported in part by
the U.S. Department of Energy, Office of Science, under Award No. DE-SC0010545.
Work at LBNL was performed under the auspices of the US DOE under
Contract DE-AC02-05CH11231 and was supported by the U.S. DOE Office
of Basic Energy Sciences, Division of Chemical Sciences.
\end{acknowledgments}

\bibliographystyle{apsrev}
\bibliography{DR}

\end{document}